\documentclass[twocolumn,showpacs,preprintnumbers,amsmath,amssymb,superscriptaddress]{revtex4}

\usepackage{graphicx}
\begin{document}
\bibliographystyle{prsty}
\title{Characterization of Magnetic Components in the Diluted Magnetic Semiconductor Zn$_{1-x}$Co$_x$O by X-ray Magnetic Circular Dichroism}

\author{M. Kobayashi}
\affiliation{Department of Physics and Department of Complexity 
Science and Engineering, University of Tokyo, 
Kashiwa, Chiba 277-8561, Japan}
\author{Y. Ishida}
\affiliation{Department of Physics and Department of Complexity 
Science and Engineering, University of Tokyo, 
Kashiwa, Chiba 277-8561, Japan}
\author{J. l. Hwang}
\affiliation{Department of Physics and Department of Complexity 
Science and Engineering, University of Tokyo, 
Kashiwa, Chiba 277-8561, Japan}
\author{T. Mizokawa}
\affiliation{Department of Physics and Department of Complexity 
Science and Engineering, University of Tokyo, 
Kashiwa, Chiba 277-8561, Japan}
\author{A. Fujimori}
\affiliation{Department of Physics and Department of Complexity 
Science and Engineering, University of Tokyo, 
Kashiwa, Chiba 277-8561, Japan}
\author{K. Mamiya}
\affiliation{Synchrotron Radiation Research Center, Japan Atomic 
Energy Research Institute, Mikazuki, 
Hyogo 679-5148, Japan}
\author{J. Okamoto}
\affiliation{Synchrotron Radiation Research Center, Japan Atomic 
Energy Research Institute, Mikazuki, 
Hyogo 679-5148, Japan}
\author{Y. Takeda}
\affiliation{Synchrotron Radiation Research Center, Japan Atomic 
Energy Research Institute, Mikazuki, 
Hyogo 679-5148, Japan}
\author{T. Okane}
\affiliation{Synchrotron Radiation Research Center, Japan Atomic 
Energy Research Institute, Mikazuki, 
Hyogo 679-5148, Japan}
\author{Y. Saitoh}
\affiliation{Synchrotron Radiation Research Center, Japan Atomic 
Energy Research Institute, Mikazuki, 
Hyogo 679-5148, Japan}
\author{Y. Muramatsu}
\affiliation{Synchrotron Radiation Research Center, Japan Atomic 
Energy Research Institute, Mikazuki, 
Hyogo 679-5148, Japan}
\author{A. Tanaka}
\affiliation{Depatment of Quantum Matter, ADSM, Hiroshima University, 
Higashi-Hiroshima 739-8530, Japan}
\author{H. Saeki}
\affiliation{Institute of Scientific and Industrial Research, 
Osaka University, Ibaraki, Osaka 567-0047, Japan}
\author{H. Tabata}
\affiliation{Institute of Scientific and Industrial Research, 
Osaka University, Ibaraki, Osaka 567-0047, Japan}
\author{T. Kawai}
\affiliation{Institute of Scientific and Industrial Research, 
Osaka University, Ibaraki, Osaka 567-0047, Japan}
\date{\today}

\begin{abstract}
We report on the results of x-ray absorption (XAS), x-ray magnetic circular dichroism (XMCD), and photoemission experiments on {\it n}-type Zn$_{1-x}$Co$_x$O ($x=0.05$) thin film, which shows ferromagnetism at room temperature. 
The XMCD spectra show a multiplet structure, characteristic of the Co$^{2+}$ ion tetrahedrally coordinated by oxygen, suggesting that the ferromagnetism comes from Co ions substituting the Zn site in ZnO. 
The magnetic field and temperature dependences of the XMCD spectra imply that the non-ferromagnetic Co ions are strongly coupled antiferromagnetically with each other.

\end{abstract}


\maketitle
Diluted magnetic semiconductors (DMS's), in which a portion of atoms of the non-magnetic semiconductor hosts are replaced by magnetic ions, are key materials for ``spintronics'' (spin electronics), which is intended to manipulate both the spin and charge degrees of freedom by use of coupling between the spins of the magnetic ions and the charge carriers of the host semiconductors \cite{SSE}. 
Indeed, using ferromagnetic DMS's, it has been successful to realize spin-related new techniques such as spin injection \cite{ESI}, electrical manipulation of magnetization reversal \cite{MagRev}, and current-induced domain-wall switching \cite{CIDWS}. 
However, because the Curie temperature ($T_{\text{C}}$) of the prototypical ferromagnetic DMS Ga$_{1-x}$Mn$_x$As is below the room temperature ($T_{\text{C}}{\textless}200$ K), it is still difficult to utilize DMS's in practical applications. 
Recently, oxide-based DMS's \cite{DMS_Oxides}, especially ZnO-based DMS's \cite{Ueda, MnZnO, ZnVO}, have attracted much attention as candidates for room temperature ferromagnetic DMS's. The wide band gap of ZnO is also expected to expand the range of applications. 
Theoretical studies have predicted that intrinsic ferromagnetism of Co-doped ZnO can be stabilized by electron doping \cite{Yoshida, Lee}. However, possible extrinsic origins of the ferromagnetism such as precipitated Co metal clusters \cite{Cluster} have not been excluded and the ferromagnetism of Zn$_{1-x}$Co$_x$O is still in strong dispute.

Although magnetization and anomalous Hall effect measurements are suitable to investigate magnetic properties, it is not straight forward to judge from these measurements whether the ferromagnetism is intrinsic or extrinsic \cite{AHE}. 
X-ray magnetic circular dichroism (XMCD), which is the difference in core-level absorption spectra between right- and left-handed circularly polarized x-rays, is an element specific probe sensitive to the magnetic polarization of each element, and therefore enables us to directly extract the local electronic structure related to particular magnetic properties of the substituted transition-metal ions \cite{GaMnAsXMCD}. 
In this work, we have performed combined x-ray absorption (XAS), XMCD and photoemission spectroscopy (PES) studies of Zn$_{1-x}$Co$_x$O to determine the electronic structure and the magnetic properties associated with the Co ions. In particular, the XMCD line shape and the intensity under varying magnetic field and temperature have implied that the ferromagnetism of Zn$_{1-x}$Co$_x$O is indeed caused by the Co$^{2+}$ ions substituting the Zn site.

A Zn$_{1-x}$Co$_x$O ($x=0.05$) thin film was epitaxially grown on a ${\alpha}$-Al$_2$O$_3$ (0001) substrate by the pulsed laser deposition technique using an ArF excimer laser with energy density 1.0 J/cm$^2$. During the deposition, the substrate was kept at a temperature of ${\sim}300$ $^{\circ}$C in an oxygen pressure of $1.0{\times}10^{-5}$ mbar. The total thickness of the Zn$_{1-x}$Co$_x$O layer was {$\sim2000$} {\AA} on a 500 {\AA} ZnO buffer layer. X-ray diffraction confirmed that the thin film had the wurtzite structure and no secondary phase was observed. Details of the sample fabrication are given in Ref.~\cite{TMSonZNO}. Ferromagnetism with $T_{\text{C}}$ above the room temperature was confirmed by magnetization measurements using a SQUID magnetometer (Quantum Design, Co. Ltd.).

XAS and XMCD measurements at the Co $2p{\to}3d$ (Co $L_{2,3}$) edge were performed at beam line BL23SU \cite{BL23SU} of SPring-8 in the total-electron yield mode. The monochromator resolution was $E/{\Delta}E\textgreater10000$. Right-handed (${\mu}^+$) and left-handed (${\mu}^-$) circularly polarized x-ray absorption spectra were obtained by reversing photon helicity at each photon energy. External magnetic field was applied perpendicular to the sample surface. In the XMCD experiment, the magnetic field ($H$) is changed from 2.0 to 7.0 T at 20 K and the temperature ($T$) from 20 to 220 K at 7.0 T. The background of the XAS spectra was assumed to be a hyperbolic tangent function as usual. Circularly polarized x-ray absorption spectra under each experimental condition have been normalized to the maximum height of the Co $L_{2, 3}$ edge XAS [$({\mu}^++{\mu}^-)/2$] spectra as 100. 
Ultraviolet photoemission (UPS) measurements were performed at BL-18A of Photon Factory (PF), High Energy Accelerator Research Organization (KEK). Spectra were taken at room temperature in a vacuum below $7.5{\times}10^{-10}$ Torr. The total resolution of the spectrometer (VG CLAM hemispherical analyzer) including temperature broadening was $\sim200$ meV. X-ray photoemission (XPS) measurements were performed using a Gammadata Scienta SES-100 hemispherical analyzer and an Al$K\alpha$ source ($h{\nu}=1486.6$ eV) in a vacuum below $1.0{\times}10^{-9}$ Torr. In both UPS and XPS measurements, photoelectrons were collected in the angle integrated mode. Sample surface was cleaned by cycles of Ar$^+$-ion sputtering at 1.5 kV and annealing at 250 $^{\circ}$C. Cleanliness of the sample surface was checked by the absence of a high binding-energy shoulder in the O $1s$ spectrum and C $1s$ contamination by XPS.

\begin{figure}[t]
\begin{center}
\includegraphics[width=9.0cm]{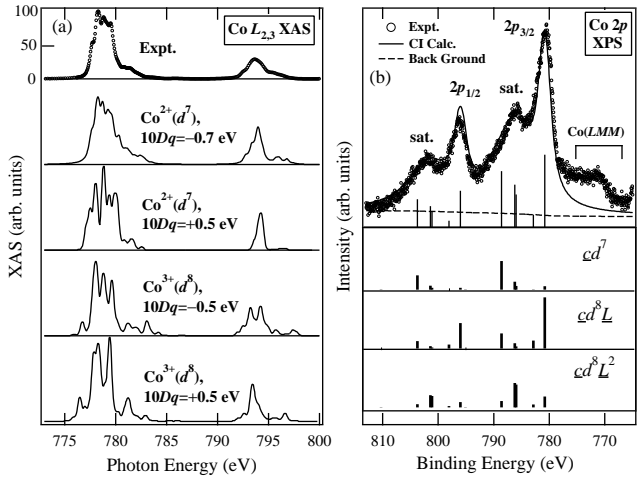}
\caption{Co $2p$ core-level spectra of Zn$_{0.95}$Co$_{0.05}$O. (a) Experimental Co $L_{2, 3}$ edge XAS spectrum (top) compared with atomic multiplet calculations, in which the Co valence and the sign and magnitude of the crystal-field 10$Dq$ are varied. The positive and negative 10$Dq$ mean that the Co ion is coordinated octahedrally and tetrahedrally by oxygen atoms. (b) Co $2p$ XPS spectrum and CI cluster-model analysis. $\underline{c}$ and $\underline{L}$ denote a hole in the Co $2p$ and the oxygen $2p$ orbitals, respectively.}
\label{CoreSpectra}
\end{center}
\end{figure}

Figure~\ref{CoreSpectra}(a) shows the Co $L_{2, 3}$ XAS spectrum compared with spectra calculated using atomic multiplet theory. The calculation was carried out for Co$^{2+}$ and Co$^{3+}$ with the positive and negative crystal-field parameter $10Dq$ representing the octahedral and tetrahedral co-ordinations of oxygen atoms for Co, respectively. 
The calculated multiplet splitting for Co$^{3+}$ is more spread than experiment both for the octahedral and tetrahedral crystal fields, and the calculated spectra for Co$^{2+}$ better reproduce the experiment. Furthermore, the negative 10$Dq$ better reproduces the measured XAS spectrum. 
Hence, we conclude that the Co ion is divalent and is tetrahedrally coordinated by four oxygen atoms.

\begin{figure}[t]
\begin{center}
\includegraphics[width=9.0cm]{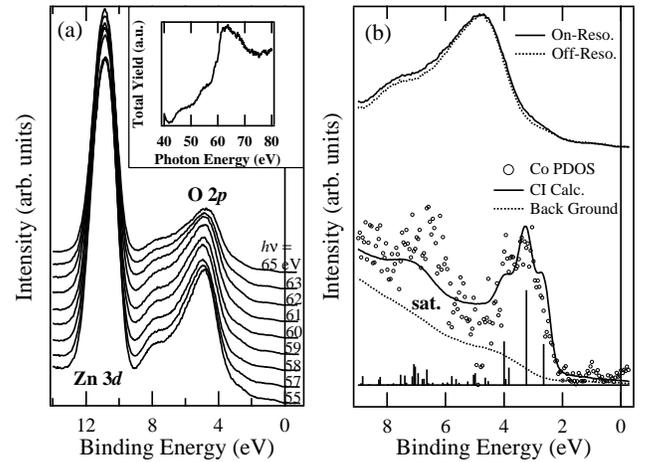}
\caption{Valence-band UPS spectra of Zn$_{0.95}$Co$_{0.05}$O. (a) A series of photoemission spectra for photon energies in the Co $3p{\to}3d$ core-excitation region. Inset: Absorption spectrum recorded in the total electron yield mode. (b) Top: On-resonance ($h{\nu}=61.5$ eV) and off-resonance ($h{\nu}=60.0$ eV) spectra. The difference between these spectra represents the Co $3d$ PDOS. Bottom: CI cluster-model analysis for the Co $3d$ PDOS.}
\label{PESVBCI}
\end{center}
\end{figure}

Further information about the local electronic structure of the Co$^{2+}$ ion, namely, hybridization of the Co $3d$ orbital with the host oxygen $2p$ orbital as well as the $d$-$d$ Coulomb interaction can be studied by photoemission spectroscopy \cite{GMA_XPS, GMA_RPES}. 
The Co core-level $2p$ XPS spectrum of Zn$_{0.95}$Co$_{0.05}$O shown in Fig.~\ref{CoreSpectra}(b) is similar to that of CoO \cite{CoO}. 
We have made a configuration-interaction (CI) cluster-model analysis of the Co $2p$ XPS spectrum using a [$\text{Co}^{2+}(\text{O}^{2-})_4]^{6-}$ cluster and estimated the electronic structure parameters: the ligand-to-$3d$ charge-transfer energy ${\Delta}=5.0{\pm}0.5$ eV, the {\it d-d} Coulomb interaction energy $U=6.0{\pm}0.5$ eV, and the Slater-Koster parameter $(pd{\sigma})=-1.6{\pm}0.1$ eV. These parameters are in good agreement with those obtained from the previous XAS study \cite{ZnCoOXAS} and are consistent with the chemical trend in transition-metal-doped II-VI DMS's \cite{CI}. 
Here, Racah parameters have been fixed at the values of the free ion: $B=0.138$ eV and $C=0.54$ eV. The Co 3{\it d-}2{\it p} core hole Coulomb attraction energy $Q$ is related to $U$ through $U={\beta}Q$, where at ${\beta}=0.7$. The ratio between $(pd{\sigma})$ and $(pd{\pi})$ has been fixed $(pd{\sigma})/(pd{\pi})=-2.16$. Figure~\ref{CoreSpectra}(b) shows that the main peak of the spectrum dominantly consists of charge-transferred states, i.e. $\underline{c}d^8{\underline{L}}$, where $\underline{c}$ and $\underline{L}$ denote a hole in the Co $2p$ and oxygen $2p$ orbitals, respectively.

Figure~\ref{PESVBCI}(a) shows the valence-band UPS spectra of Zn$_{0.95}$Co$_{0.05}$O taken at various photon energies in the Co $3p{\to}3d$ core-excitation region. The absorption spectrum in the same energy region is shown in the inset. Binding energies ($E_{\text{B}}$'s) are referenced to the Fermi level ($E_{\text{F}}$) of a metallic sample holder which is in electrical contact with the sample. The absorption spectrum shows that Co $3p{\to}3d$ absorption occurs at $h{\nu}{\sim}61$ eV. Constant-initial-state (CIS) spectra at various $E_{\text{B}}$'s (not shown) indicate that the Co $3d$ partial density of states (PDOS) is primarily located at $E_{\text{B}}{\sim}3.0$ and $\sim7.0$ eV. 
Figure~\ref{PESVBCI}(b) shows the Co $3d$ PDOS of Zn$_{0.95}$Co$_{0.05}$O, which has been obtained by subtracting the off-resonance ($h{\nu}=60$ eV) spectrum from the on-resonance ($h{\nu}=61.5$ eV) one. Here, the off-resonance spectrum was multiplied by the integrated ($0{\textless}E_{\text{B}}{\textless}9$ eV) intensity ratio between $h{\nu}=61.5$ and 60 eV of pure ZnO. The Co $3d$ PDOS shows a peak at $E_{\text{B}}{\sim}3.0$ eV, which is similar to that of the polycrystalline Zn$_{0.9}$Co$_{0.1}$O \cite{ZnCoOPES}, and a satellite at $E_{\text{B}}{\sim}7.0$ eV. Because the energy difference between the top of the O $2p$ band and $E_{\text{F}}$ is nearly equal to the band gap of ZnO, $E_{\text{F}}$ is supposed to be located near the conduction band minimum (composed of Zn $4s$ and possibly of Co $d$ states), meaning that the sample is {\it n}-type. Although local-density approximation (LDA) calculations have predicted that ferromagnetism is mediated by carriers and therefore needs a high density of states (DOS) at $E_{\text{F}}$ \cite{Yoshida}, we could not clearly observe a finite DOS at $E_{\text{F}}$, consistent with the low carrier density. 
The energy difference between the main structure and the satellite of the Co $3d$ PDOS was as large ${\sim}9$ eV for CoO \cite{CoO} while it is $\sim$4 eV for Zn$_{1-x}$Co$_x$O, probably because of the different co-ordinations of oxygen atoms between CoO and Zn$_{1-x}$Co$_x$O. 
This can be well explained by the CI cluster-model calculation using the same ${\Delta}$, $U$, $(pd{\sigma})$ as shown in Fig.~\ref{PESVBCI}(b).

\begin{figure}[t]
\begin{center}
\includegraphics[width=9.0cm]{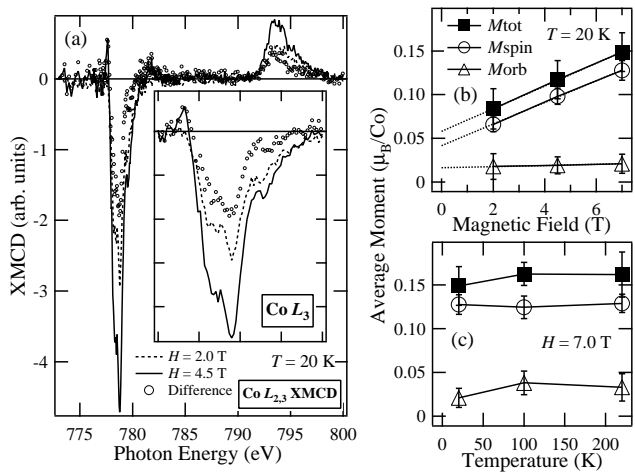}
\caption{
Magnetic field and temperature dependences of the Co $L_{2, 3}$ XMCD spectra of Zn$_{0.95}$Co$_{0.05}$O. 
(a) XMCD spectra under different magnetic fields at 20 K. Open circle shows the difference of the spectra between $H=4.5$ and 2.0 T. 
(b) Average magnetic moments $M_{\text{spin}}$, $M_{\text{orb}}$, and $M_{\text{tot}}$ as functions of magnetic field, estimated using the XMCD sum rules. (c) The same as (b) as function of temperature. 
}
\label{XMCD}
\end{center}
\end{figure}

Although the XAS spectra are independent of the magnetic field and ${\mu}^+$ and ${\mu}^-$ are nearly identical on the scale of Fig.~\ref{CoreSpectra}(a), there were weak but reprodusible XMCD signals (${\mu}^+-{\mu}^-$) as shown in Fig.~\ref{XMCD}(a). 
The intensity of XMCD spectra increases with increasing magnetic field as shown in Fig.~\ref{XMCD}(b), while it is rather independent of temperature as shown in Fig.~\ref{XMCD}(c). 
Note that, in Fig.~\ref{XMCD}(b) and (c), the spin ($M_{\text{spin}}$), orbital ($M_{\text{orb}}$), and total ($M_{\text{tot}}$) magnetic moments of Co estimated using XMCD sum rules \cite{orbSR, spinSR} are plotted rather than the raw XMCD intensities. 
Part of the XMCD signals which linearly increases with $H$ represents the paramagnetic component, while XMCD signals which persist at $H{\sim}0$ T represent the ferromagnetic component \cite{FMPM}. 
The difference between XMCD spectra under $H=2.0$ and 4.5 T, which reflects the paramagnetic component, shows nearly the same line shape as the XMCD spectrum taken at the lowest magnetic field of 2.0 T, as shown in Fig.~\ref{XMCD}(a). Therefore, it seems that the Co ions have similar electronic structures in the paramagnetic and ferromagnetic components. It should be emphasized that the XMCD spectra also show a multiplet structure, unlike those of Co metal \cite{Chen}, indicating that the magnetism in the present sample is not due to metallic Co clusters but due to Co ions with localized $3d$ electrons.

\begin{figure}[t]
\begin{center}
\includegraphics[width=9.0cm]{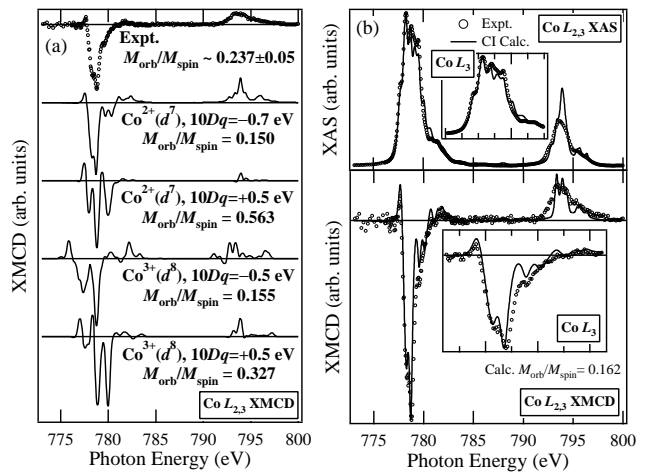}
\caption{Co $L_{2, 3}$ XMCD spectra of Zn$_{0.95}$Co$_{0.05}$O. (a) Comparison with atomic multiplet calculation, in which the valence of Co and the sign and magnitude of the crystal-field splitting are varied. (b) CI cluster-model analysis for the XAS and XMCD spectra.}
\label{XMCD_Calc}
\end{center}
\end{figure}

As in the case of the XAS spectrum, the XMCD spectra are also compared with theoretical XMCD spectra calculated using atomic multiplet theory as shown in Fig.~\ref{XMCD_Calc}(a). 
The calculated ratio $M_{\text{orb}}/M_{\text{spin}}$ is compared with experiment. 
The line shape of the calculated spectra for Co$^{3+}$ are different from the experimental one and the line shape for Co$^{2+}$ with tetrahedral oxygen co-ordination best agrees with experiment. Comparison of the ratio $M_{\text{orb}}/M_{\text{spin}}$ between the calculated spectra and experiment indicates that the Co$^{2+}$ ion with tetrahedral oxygen co-ordination and not octahedral one is consistent with experiment. 
Therefore, the calculated spectrum for Co$^{2+}$ with $10Dq=-0.7$ eV best reproduces the experimental XMCD. 
To examine more details of the XAS and XMCD line shapes, CI cluster-model calculations were performed as shown in Fig.~\ref{XMCD_Calc}(b). 
We have used electronic structure parameters ${\Delta}=5.0$ eV, $U=5.0$ eV, and $(pd{\sigma})=-1.6$ eV, nearly the same as those estimated from the PES experiments. 
The calculated $M_{\text{orb}}/M_{\text{spin}}$ is closer to the experimental value than that of atomic multiplet theory. 
This gives further support that the substituted Co ions under tetrahedral crystal field are responsible for the ferromagnetism in Zn$_{1-x}$Co$_x$O.

Finally we comment on possible origins of the temperature-independent paramagnetic component of the XMCD signals. 
This signal cannot be due to Pauli paramagnetism of conduction electrons because their XMCD spectra show characteristic of the Co$^{2+}$ ion and also that its susceptibility $\chi_{\text{exp}} \sim 1.43 \times 10^{-2}$ ($\mu_{\text{B}} \text{/T} \, \text{per\, Co}$) is several orders of magnitude larger than the Pauli paramagnetism expected for the conduction electron concentration $n \sim 1.0 \times 10^{17}$ cm$^{-3}$ of the present sample. 
If the temperature-independent paramagnetism is due to the Co$^{2+}$ ions, there should be strong antiferromagnetic interaction between the Co ions because, if the Co ions do not interact with each other, they would show the Curie behavior. 
The temperature-independent paramagnetic component may arise from the susceptibility of the antiferromagnetic Co ions having a N\'{e}el temperature above the room temperature. Such magnetic susceptibility is estimated to be $g^2S(S+1)/T_{\mathrm{N}} \sim 8.4 \times 10^{-3}$ $(\mu_{\mathrm{B}}/\mathrm{T}\,\mathrm{per\,Co})$ for $g=2$, $S=3/2$, and $T_{\mathrm{N}}=400$ K, in reasonable agreement with slope of the XMCD intensity [Fig.~\ref{XMCD}(b)] of $\chi_{\text{exp}} \sim 1.43 \times 10^{-2}$ ($\mu_{\text{B}} \text{/T} \, \text{per\, Co}$). 
Although both the ferromagnetic and paramagnetic/antiferromagnetic Co ions has the same 2+ valence and the tetrahedral crystal field, subtle differences such as neighboring defects and local lattice distortion may have lead to the different magnetic behaviors. 
In order to confirm the above conjectures, more precise and systematic XMCD measurements on samples with varying carrier concentrations are necessary.

In summary, we have performed XAS, XMCD, and PES experiments on the diluted ferromagnetic semiconductor {\it n}-type Zn$_{1-x}$Co$_x$O ($x=0.05$). 
The XMCD spectra show a multiplet structure, characteristic of the Co$^{2+}$ ion tetrahedrally coordinated by oxygen. This implies that the ferromagnetism in Zn$_{1-x}$Co$_x$O is caused by the substituted Co$^{2+}$ ions at the Zn site. The magnetic field and temperature dependences of the XMCD intensity suggest that 
the non-ferromagnetic Co ions are strongly coupled antiferromagnetically with each other.

We thank T. Okuda, A. Harasawa, and T. Kinoshita for technical help at PF. We also thank the Materials Design and Characterization Laboratory, Institute for Solid State Physics, University of Tokyo, for the use of the SQUID magnetometer. This work was supported by a Grant-in-Aid for Scientific Research in Priority Area ``Semiconductor Nanospintronics'' (14076209) from the Ministry of Education, Culture, Sports, Science and Technology, Japan. The experiment at PF was approved by the Photon Factory Program Advisory Committee (Proposal No. 2002G027). MK acknowledges support from the Japan Society for the Promotion of Science for Young Scientists.

\end{document}